\documentstyle[aps]{revtex}
\begin{document}
\begin{title}
{\bf Edge Magnetoplasmons 
For Very Low Temperatures And Sharp Density Profiles } 
\end{title}
\author{O. G. Balev$^{*}$ and P. Vasilopoulos$^\dagger$} 
\address{$^{*}$Institute of Physics of Semiconductors, National 
Academy of Sciences,
45 Pr. Nauky, Kiev 252650, Ukraine
\\
$\ ^\dagger$Concordia University, Department of Physics, 1455 de 
Maisonneuve Blvd O, Montr\'{e}al, Qu\'{e}bec, Canada, H3G 1M8}
\date{May 2, 1997}
\maketitle
\begin{abstract}
A treatment of edge magnetoplasmons (EMP), based on a {\it microscopic} evaluation of the local contributions to the current density, is presented. It is valid in the quantum Hall regime for
filling factor $\nu=1$ or $2$ and low temperatures when the dissipation is 
localized near the edge.  The confining potential, flat in the 
interior of the channel, is assumed smooth 
on the magnetic length $\ell_{0}$ scale but sufficiently steep 
at the edges that the density profile is sharp and
the dissipation considered results only from electron {\it 
intraedge-intralevel} transitions due to scattering by piezoelectrical 
phonons. For wide channels there exist independent EMP modes spatially symmetric and 
antisymmetric with respect to the edge. Certain of these modes can propagate nearly undamoped even when the dissipation is strong and are thus termed {\it edge helicons}. In contrast with well-known 
results for a spatially homogeneous dissipation within the channel, we 
obtain that the damping of the fundamental EMP  is not 
quantized and varies as $T^{3}$ or $T^{-3}$, where $T$ is the 
temperature, in the high- and low-frequency limits, respectively.  
The  characteristic length of the resulting dispersion 
relation and of the charge density distortion is
$\ell_{0}$.  The screening of the metallic gates, when present, is 
taken into account. 
\ \\
PACS\ \ 73.20.Dx, 73.40.Hm
\end{abstract}
\section{INTRODUCTION.}

In the past few years there has been considerable interest in edge 
magnetoplasmons (EMP) as well as in other edge excitations of 
two-dimensional (2D) electron systems in the presence of a magnetic 
field $B$ \cite{1}-\cite{12}. For a 2D system with a vertical 
conductivity drop at the boundaries, it has been shown \cite{1} that 
the dissipation can determine the EMP dispersion relation and the 
spatial structure in an essential manner even in the regime of the 
quantum Hall effect (QHE). In this work the properties of the EMP are 
expressed in terms of the components of the magnetoconductivity 
tensor of an infinite 2D system. Moreover, due the very low frequency 
$\omega$ of the EMP, the dispersion relation could be written in 
terms of the static magnetoconductivity tensor. 

The distance of the ``center of gravity'' of the EMP charge from the 
edge, 
which coincides with the characteristic length over which the 
transverse to the edge electric field 
$E_{y}$ of the EMP decreases, is given, for $|k_{x}\ell_{v}|\ll 1$, by 
\cite{1}

\begin{equation}
\ell_{vc}=\frac{|\ell_{v}|}{\pi}\ [ln (\frac{2}{|k_{x}\ell_{v}|})+1]= 
|\frac{\sigma_{yy}}{k_{x} \sigma_{xy}}|. 
\label{1}
\end{equation}
Here $\sigma_{yy}$ and $\sigma_{yx}$ are the conductivity components
of an infinite 2D system, 
$k_{x}$ is the EMP wave vector, and $\ell_{v}$ denotes a 
characteristic length determined by Eq.  (10) of Ref.  \cite{1}.  In 
the QHE regime for typically observed \cite{7} EMP, Eq.  (\ref{1}) 
gives $\ell_{vc} \alt 1\mu$m.  Now, in Ref.  \cite{13} we have shown 
theoretically, and in agreement with experimental observations, that 
in the QHE regime for sufficiently smooth confinement the dissipation 
is due to  intralevel-intraedge transitions of electrons 
scattered by piezoelectrical phonons and occurs mainly near the edges 
of channels.  In the linear response regime this is the main dissipation 
for channels of width $W\alt 100\ \mu$m and temperatures $T \alt 1$K
if the group velocity of edge states, $v_{g}$, is larger than the 
speed of sound $s$. 
As for dissipation in the bulk, it is exponentially suppressed for 
$T\rightarrow 0$.  Given that and the fact that the dissipation, when 
deriving Eq.  (\ref{1}), occurs in the bulk, we expect the properties 
of the EMP to be strongly modified when the dissipation is localized 
near the edges.  

The above  expectation is further supported by the results
of Ref.\cite{11} which pertain to EMP  for a smooth,
unperturbed electron density
profile which  contrasts sharply with that of Ref. \cite{1} where the density
drops vertically at the edges. In addition to the modes of Ref.\cite{1}
acoustic EMPs were obtained in Ref.\cite{11}. Further, our results 
despite
their partial similarity with those of Refs. \cite{1} and \cite{11},  show significant 
differences from them even  when the dissipation is very weak. 
For the very low temperatures that we
consider here, $k_{B}T\ll \hbar v_{g}/\ell_{0}$, and the assumed smooth
confining potential  on the scale of $\ell_{0}$ ($v_{g}>s$), the
unperturbed electron density $n_{0}(y)$, normalized to the bulk value 
$n_{0}$, drops essentially, on the scale of $\ell_{0}$, only near the edge.
More explicitly, for the potential that is specified at the beginning of Sec.
II, we calculate $n_{0}(y)/n_{0}=\{1+\Phi[(y_{re}-y)/\ell_{0}]\}/2$, where $y_{re}$ is 
the coordinate of the right edge and $\Phi(y)$ the probability integral.
In Fig. 1 we show this calculated density profile (short-dashed curve) together with 
those assumed in  Refs. \cite{1} (solid curve) and \cite{11} (long-dashed curve). 
The profile of Ref. \cite{11} is obtained with  $n_{0}(y)/n_{0}=
(2/\pi) arctan\sqrt{(y_{re}-y)/a}$ and $a/\ell_{0}=20$; it corresponds 
approximately to $a=2000\AA$. As can be seen, the three density 
profiles are very different from each other. As will be shown in this
paper, combining our density profile  with the localization of the dissipation 
near the edge leads to strong modications of the EMPs results. 

These modifications, as well as  new EMP resulting from the 
microscopic treatment of the problem, are the subject of this work.  
The description of the inhomogeneous current density in the 
quasi-static regime is carried out using the results of Ref.  
\cite{14}.  We consider the QHE regime, mainly $\nu=1$ and partly 
$\nu=2$, for samples with sufficiently large in-plane dimensions, as 
is typical in EMP experiments, that inter-edge electron transitions 
and the inter-edge Coulomb interaction can be neglected.

In Sec. II we start with the expressions for the inhomogeneous current 
densities and conductivities and derive the integral equation for EMP. 
In Sec. III we derive the dispersion relation for very low temperatures 
and in Sec. IV we describe in detail the new edge waves. Finally, in 
Sec. V we compare our theory with the experiment  and make concluding remarks.

\section{BASIC RELATIONS}

\subsection{Inhomogeneous current density in quasi-static regime} 

We consider a two-dimensional electron gas (2DEG), of width $W$, 
of length $L_x=L$, and of thickness zero, in the presence of a strong 
magnetic field $B$ parallel to the $z$ axis.
The 2DEG is confined along the $y$ axis.  For simplicity we take the 
confining potential as parabolic at the edges: $V_{y}^{'}=0$, for 
$y_l<y<y_r$, $V_{y}^{'}=m^*\Omega^2 (y-y_r)^2/2$ for $y>y_r>0$, and 
$V_{y}^{'}=m^* \Omega^2 (y-y_l)^2/2$ for $y<y_l<0$.  Because in real 
EMP experiments $W \agt 0.1 $cm, we can assume $[W-(y_r-y_l)]/W \ll 
1$.  Moreover, we will assume that $|k_{x}| W \gg 1$ such that it is 
possible to consider an EMP along the right edge of the channel, of 
the form $A(\omega, k_{x}, y) \exp[-i(\omega t-k_{x} x)]$, totally 
independent of the left edge.  We consider only linear responses.
For definiteness, we take the background dielectric constant $\epsilon$ 
to be spatially homogeneous.  We consider $B$ strong enough that only 
the $n=0$ Landau level (LL) is occupied.  For the $\nu=2$ QHE regime 
we will neglect the spin-splitting.  As for the $\nu=1$ QHE, we will 
assume that the spin-splitting, caused by many-body effects, is strong 
enough to neglect the contribution related with the upper spin-split 
LL. We assume a lateral confinement smooth on the scale of the 
magnetic length $\ell_{0}=(\hbar/m^{*} \omega_{c})^{1/2}$ such that 
$\Omega \ll \omega_{c}$, where $\omega_{c}=|e|B/m^* $ is the cyclotron 
frequency.  Further, we will approximate 
$\tilde{\omega}=(\omega_{c}^{2}+\Omega^2)^{1/2}$ by $\omega_{c}$.

Because the EMP is practically quasi-static and its wavelength
$\lambda \agt 1$cm is 
very large, we expect, in analogy with well-known results that 
follow from Maxwell's equations \cite{15}, the associated electric field 
 $E_{x}(x,y,t)$ to have a smooth dependence on $y$ on the scale of 
$\ell_{0}=(\hbar/m^{*} \omega_{c})^{1/2}$, i. e.,
$E_{x}(x,y,t) = E_{x}(y) \exp[-i(\omega t-k_{x} x)]$. Physically it is clear: 
the dependence of $E_{x}(x,y,t)$ on $y$, as expressed through
Maxwell's equations, is  related to
that on $x$ which has a characteristic scale $\lambda$. Thus, $E_{x}(y)$ 
should have the same scale $\lambda$ and be definitely smooth on the $\ell_{0}$ 
scale. This is a general result and applies to
 the case treated in Ref. \cite{1}. Then using the results of Ref. 
\cite{14}, we obtain the components of the current density in the form
		
		\begin{equation}
		j_y(y)=\sigma_{yy}(y) E_{y}(y)+\sigma_{yx}^0(y) E_{x}(y),\\ 
		\label{2}
		\end{equation}
		
		\begin{equation}
		j_x(y)=\sigma_{xx}(y)E_x(y)-\sigma_{yx}^0(y) E_{y}(y)+ 
		v_{g}\rho(\omega, k_{x},y). 
		\label{3}
		\end{equation}
Here  we have suppressed the exponential factor $\exp[-i(\omega 
t-k_{x} x)]$ common to all terms in Eqs. (\ref{2}) and (\ref{3}). It is
understood that $E_{\mu}(y)$ depends on $\omega$ and $k_{x}$. 
As follows from Refs.  \cite{13} and \cite{14}, $\sigma_{yy}(y)$
is strongly (exponentially) 
localized at the edge, within a distance $\alt \ell_{0}$ from it, for  
$\hbar v_{g}\gg k_{B}T\ell_{0}$. The last term on the riht-hand side 
(RHS) of Eq. (\ref{3}), absent in Ref. \cite{14}, represents a 
convection contribution to the current density along $x$, associated 
with the wave, and is due to a distortion $\delta\rho$ of the charge 
localized near the edge; we denote it by $\rho(\omega, k_{x},y)$ in 
order to simplicify the notation as it occurs frequently. 
Notice that in Ref. \cite{14} the contributions to the components of
the current density are microscopically obtained for the electric
field components smooth on the scale of $\ell_{0}$. This condition holds
for the contributions $\propto E_x(y)$ in Eqs. (\ref{2}) and (\ref{3}) but is not well justified for those $\propto E_y(y)$. We assume that the latter
can be reasonably  approximated by those obtained microscopically when $E_y(y)$ is smooth on the scale of $ \ell_{0}$. The assumption is equivalent to  neglecting possible nonlocal   contributions to the current density
$\propto \int dy' \sigma_{\mu y}(y,y') E_{y}(y')$; then  it follows  that $\sigma{xy}^{0}=-\sigma{yx}^{0}$. For $\nu=1$  
we have \cite{14} 

		\begin{equation}
		\sigma_{yx}^0(y)=\frac{e^2}{2\pi\hbar}
		\int_{-\infty}^\infty dy_{0\alpha}f_{\alpha 0} \Psi_{0}^2(y-y_{0 
		\alpha}),
		\label{4}
		\end{equation}
where $\alpha\equiv \{0,k_{x\alpha}\}$,\ 
$y_{0 \alpha}=\ell_{0}^{2} k_{x\alpha}$, $\Psi_{n}(y)$ is a harmonic 
oscillator function, and $f_{\alpha 0} \equiv f_{0}(k_{x\alpha})= 
1/[1+exp((E_{\alpha 0}-E_{F0})/k_BT)]$ is the Fermi-Dirac function.
 $E_{F0}$ is the Fermi level counted from the bottom of the lowest 
 electric subband. For $T=0$ and near the right edge we have 
$\sigma_{yx}^{0}(y)=
(e^{2}/4\pi\hbar)\{1+\Phi[(y_{re}-y)/\ell_{0}]\}$, where $\Phi(x)$ is the 
probability integral, $y_{re}=\ell_{0}^{2}k_{re}$, and $f_{0}(k_{re})=1/2$. 
That is, $\sigma_{yx}^{0}(y)$ near the edge decreases on the scale of $\ell_{0}$
and behaves as the density of the short-dashed curve of Fig. 1.
Considering only the right edge 
and the flat part of the confining potential, for $y_l \leq y_{0 
\alpha} \leq y_r$ we have $E_{\alpha 0}=\hbar \omega_{c}/2$ and for 
$y_{0 \alpha} \geq y_r$ we obtain  

		\begin{equation}
		E_{\alpha 0} \equiv E_{0}(k_{x\alpha})=
		\hbar \omega_{c}/2+m^*\Omega^2 (y_{0 \alpha}-y_r)^2/2 . 
		\label{5}
		\end{equation}
We consider only the interaction of electrons with phonons, and 
neglect that with impurities, since the former is
the most essential for the assumed conditions \cite{13}. Because of the very 
smooth dependence of $E_x(y)$ on  $\ell_{0}$, we can assume that 
$\sigma_{xx}(y)$ can be  approximated by $\sigma_{yy}(y)$
which follows from Eq. (16) of Ref. \cite{14} as 
		\begin{eqnarray}
		\nonumber
		\sigma_{yy}(y)=&&\frac{\pi e^{2}\ell_{0}^{4}}{4\hbar Lk_{B}T}
		\sum_{k_{x \alpha} {\bf q}} |C_{{\bf q}}|^2 q_{x}^{2}
		[f_{0}(k_{x\alpha}-q_{x})-f_{0}(k_{x\alpha})]
		\delta [E_{0}(k_{x\alpha}) - E_0(k_{x\alpha}-q_{x})-\hbar\omega_{\vec q}]\\* 
		&&\times 
		e^{-(q_{x}^{2}+q_{y}^{2})\ell_{0}^{2}/2}\ sinh^{-2}
		(\frac{\hbar\omega_{\vec q}}{2k_{B}T})
		\ [\Psi_{0}^2(y-y_0(k_{x\alpha}-q_{x})) 
		+\Psi_{0}^2(y-y_0(k_{x\alpha}))].
		\label{6}
		\end{eqnarray}
For the low temperatures pertinent to the quantum Hall effect we consider 
only the standard acoustical (DA-) or piezoelectrical (PA-) phonons for which 
$\omega_{{\bf q}}=s q$, where $s$ is the speed of sound, and
$q=\sqrt{q_{x}^{2}+q_{y}^{2}+q_{z}^{2}}$. Then 
$|C_{{\bf q}}|^2=(c^{'}/L_xL_yL_z)q^{\pm 1}$, where $+1$ is for DA- 
and $-1$ for PA-phonons, respectively.

\subsection{Integral equation for EMP with dissipation at the edges} 

Using Eqs. (\ref{2})-(\ref{4}), and (\ref{6}), we can write the 
continuity equation, linearized in $\delta\rho(\omega, k_{x},y)\equiv
\rho(\omega, k_{x},y)$, as
		\begin{eqnarray}
		\nonumber
		-i(\omega-k_{x}v_{g})\rho(\omega, k_{x},y)+&&
		ik_{x}[\sigma_{xx}(y)\ E_{x}(\omega, k_{x},y)-
		\sigma_{yx}^{0}(y) E_{y}(\omega, k_{x},y)]\\*
		&&+\frac{\partial}{\partial y}[\sigma_{yy}(y) E_{y}(\omega, k_{x},y)
		+\sigma_{yx}^{0}(y) E_{x}(\omega, k_{x},y)]=0.
		\label{7}
		\end{eqnarray}

In terms of the potential $\phi(\omega, k_{x},y)$ the electric field 
components are $ E_{x}(\omega, k_{x},y)=-ik_{x}\phi(\omega, k_{x},y)$ 
and $ E_{y}(\omega, k_{x},y)= -\frac{\partial}{\partial y} 
\phi(\omega, k_{x},y)$.  Then Eq.  (\ref{7}) gives
		\begin{eqnarray}
		\nonumber
		-i(\omega-k_{x}v_{g})\rho(\omega, k_{x},y)&&+k_{x}^{2}\sigma_{xx}(y)\phi(\omega, k_{x},y)\\*
		&&-\frac{\partial}{\partial y}[\sigma_{yy} (y) \frac{\partial}{\partial y}
		\phi(\omega, k_{x},y)]
		-ik_{x}\phi(\omega, k_{x},y)\frac{d}{dy}\sigma_{yx}^{0}(y)=0. 
		\label{9}
		\end{eqnarray}
Now using Poisson's equation we obtain
		\begin{equation}
		\phi(\omega, k_{x},y)=\frac{2}{\epsilon}\int_{-\infty}^{\infty}
		dy' K_{0}(|k_{x}||y-y'|)\rho(\omega, k_{x},y'),
		\label{10}
		\end{equation}
where $ K_{0}(x)$ is the mofified Bessel function; $\phi$ and 
$\rho$ pertain to the 2D plane. From Eqs. (\ref{9}) and (\ref{10}) we 
obtain the following integral equation for  $\rho(\omega, k_{x},y)$
		\begin{eqnarray}
		\nonumber
		-i(\omega-k_{x}v_{g})&&\rho(\omega, k_{x},y)+\frac{2}{\epsilon}
		\{ k_{x}^{2}\sigma_{xx}(y)
		-ik_{x}\frac{d}{d y}[\sigma_{yx}^{0}(y)]\\*
		\nonumber
		\ \\
		&&-\sigma_{yy}(y)\frac{d^{2}}{d y^{2}}
		-\frac{d}{d y}[\sigma_{yy}(y)]\frac{d}{d y}\}
		\int_{-\infty}^{\infty}
		dy' K_{0}(|k_{x}||y-y'|)\rho(\omega, k_{x},y')=0. 
		\label{11}
		\end{eqnarray}
The value of $\sigma_{yy}(y)$  is
significantly different than zero only near the edges of the channel. 
The same holds for the values of $\sigma_{xx}(y)$ and of
$d\sigma_{yx}^{0}(y)/d y$; for $\hbar v_{g}\gg \ell_{0}k_{B}T$
 this can be seen
from Eqs. (\ref{4}) and (\ref{6}) which show that  $\sigma_{yy}(y)$ and
$d\sigma_{yx}^{0}(y)/d y$ are exponentially localized within
a distance $\approx \ell_{0}$ from the right edge at $y_{re}=y_{r}+\Delta y_{r}$. 
We have $\Delta 
y_{r}=\ell_{0}^{2}k_{e}$ where $k_{e}=(\omega_{c}/ \hbar\Omega)
\sqrt{2m^{*}\Delta_{F}}$ is the characteristic wave 
vector associated with an edge state, $\Delta_{F}=E_{F0}-\hbar 
\omega_{c}/2$,
and $W=2y_{re}$. 
For $k_{x\alpha}\equiv k_{re}=
y_{r}/\ell_{0}^{2}+ k_{e}$ we have $f_{0}(k_{re})=1/2$ and 

\begin{equation}
v_{g}=\frac{1}{\hbar}\frac{\partial E_{0}(k_{re})}{\partial 
k_{x\alpha}}=\frac{\hbar 
\Omega^{2}k_{e}}{m^{*}\omega_{c}^{2}}=\sqrt{\frac{2\Delta_{F}}{m^{*}}}
\frac{\Omega}{\omega_{c}}.
\label{12}
\end{equation}
Eq. (\ref{12}) can also be written as $v_{g}=E_{e}/B$, where 
$E_{e}=\Omega \sqrt{2m^{*}\Delta_{F}}/|e|$ is the electric field 
describing the influence of the confining potential.
For a dissipationless, 2D classical electron liquid we have, for 
finite $\omega$, 
$\sigma_{yy}(y)=\sigma_{xx}(y)=ie^{2}n_{0}(y)\omega/m^{*}
(\omega^{2}-\omega_{c}^{2})$ and $\sigma_{yx}^{0}(y) 
=-e^{2}n_{0}(y)\omega_{c}/m^{*}
(\omega^{2}-\omega_{c}^{2})$, where $n_{0}$ is the electron density. 
Then Eq. (\ref{11}) becomes identical 
with Eq. (4) of Ref. \cite{11}. In addition, if we 
assume that the conductivity components in Eq. (\ref{11}) are 
independent of $y$, for $|y|<W/2$, and $\sigma_{yy}(y)=\sigma_{xx}(y)$, Eq. (\ref{11})
takes the form of Eq. (15) of Ref. \cite{1} after integration over $z$. 

Equations (\ref{10}) and (\ref{11})  apply to a 2DEG in the 
absence of metallic gates.  Sometimes a metallic gate is placed on the 
top of the sample \cite{10} at a distance $d$ from the 
2DEG. As shown in the Appendix, for a gated sample the kernel $K_{0}$ in 
Eqs.  (\ref{10}) and (\ref{11}) is replaced by $R_{g}=K_{0}(|k_{x}||y-y'|)- 
K_{0}(|k_{x}| \sqrt{(y-y')^{2}+4d^{2}})$. If this gate is replaced by 
air, then $K_{0}$ is replaced by $R_{a}=K_{0}(|k_{x}||y-y'|)+ 
[(\epsilon -1)/(\epsilon +1)]K_{0}(|k_{x}| \sqrt{(y-y')^{2}+4d^{2}})$.

\section{EMP DISPERSION RELATION}


We  consider very low temperatures that satisfy the inequality $\hbar 
v_{g}\gg \ell_{0}k_{B}T$. From Eqs.  (\ref{4}) and  
(\ref{6}) it  follows that $d\sigma_{yx}^{0}(y)/d y
=-(e^{2}/2\pi\hbar)\Psi_{0}^{2}( y-y_{re})$; also, $\sigma_{yy}(y)$  and 
$\sigma_{xx}(y)$ behave as $\Psi_{0}^2(y-y_{re})$ and hence 
are strongly concentrated near the edge. It follows from Eq.  
(\ref{11}) that $\rho(\omega, k_{x},y)$ is also concentrated 
near the edge.  Integrating Eq.  (\ref{11}) over $y$, from 
$y_{re}-\Delta y$ to $y_{re}+\Delta y$ with $\Delta y\sim\ell_{0}$, we 
obtain 
	
		\begin{equation}
		\int_{-\infty}^{\infty}dy\rho(\omega, 
		k_{x},y) [-(\omega-k_{x}v_{g})
		+S\ \int_{-\infty}^{\infty}dy' 
		\Psi_{0}^2(y'-y_{re})
		K_{0}(|k_{x}||y-y'|)]=0; 
		\label{13}
		\end{equation}
here $S=(2/\epsilon)(-i k_{x}^{2}\tilde{\sigma}_{xx} 
+k_{x}\sigma_{yx}^{0})$, $ \sigma_{yx}^{0}= e^{2}/2\pi\hbar$ is the 
Hall conductivity in the bulk as follows from Eq.  (\ref{4}), and 
$\tilde{\sigma}_{\mu\mu} 
=\sigma_{\mu\mu}(y)/\Psi_{0}^2(y-y_0(k_{re}))$, $\mu=x,y$.  
Using Eq.  (\ref{6}) for $v_{g}>s$ we obtain
		\begin{equation}
		\tilde{\sigma}_{xx}\approx \tilde{\sigma}_{yy}=
		 \frac{3 e^{2}\ell_{0}^{4}c'k_{B}^{3}T^{3}}{2\pi^{2}\hbar^{6} 
		v_{g}^{4}s}.
		\label{14}
		\end{equation}
Eq.  (\ref{14}) coincides with $j_{d}W/E_{y}$, given by Eq.  (32) of 
Ref.  \cite{13}a, for $E_{y}\rightarrow 0$.
 For $v_{g}<s$ the contribution to $\tilde{\sigma}_{xx}$ is exponentially 
suppressed and has an activated behavior \cite{13}.  For $v_{g}^{2}\gg s^{2}$ 
we have $\tilde{\sigma}_{xx}=e^{2}\ell_{0}^{3}c'k_{B}^{2}T^{2}/\sqrt{2}\pi^{5/2}
\hbar^{5}v_{g}^{4}$, if   $1\gg k_{B}T\ell_{0}/\hbar 
v_{g}>s/\sqrt{2}v_{g}$, and Eq. (\ref{14}) if $1\gg k_{B}T\ell_{0}/\hbar v_{g}
<s/\sqrt{2}v_{g}$  .

Notice that the terms  in Eq. 
(\ref{11}) related to $\sigma_{yy}(y)$ are totally absent in  Eq. (\ref{13}). Now, 
 in Eq. (\ref{13}) we have $|y-y'|\sim\ell_{0}$ and 
 $|k_{x}|\ell_{0}\sim 10^{-6}$; then we 
can  make the approximation $K_{0}(|k_{x}||y-y'|)\approx ln 
(2/|k_{x}\ell_{0}|)-\gamma-ln (|y-y'|/\ell_{0})$ where $\gamma$ is 
the Euler constant. The value of the integral over $y'$ in  Eq. (\ref{13}) is 
$ln (2/|k_{x}|\ell_{0})-\gamma-(1/\sqrt{\pi})\int_{-\infty}^{\infty}dt 
e^{-t^{2}}ln |t-(y-y_{re})/\ell_{0}|$. For the gated sample and that 
with air above $z=d$, $4d^{2}\gg \ell_{0}^{2}$,
the corresponding approximations in the long wavelength limit 
are  $R_{g}\approx 
ln(2d/\ell_{0})-ln(|y-y^{`}|/\ell_{0})$ and, with $\epsilon\gg 1$,  
  $R_{a}\approx 
ln(2/k_{x}^{2}d\ell_{0}) -2\gamma -ln(|y-y^{`}|/\ell_{0})$, respectively.

Now for $T\rightarrow 0$ we have $\tilde{\sigma}_{xx} \rightarrow 0$,
$\tilde{\sigma}_{yy} \rightarrow 0$ and  Eq. (\ref{11}) shows that   
$\rho(\omega,k_{x},y)$ behaves essentially as 
$\Psi_{0}^{2}(y-y_{re})$. Then the value related to the integral over $t$ can be 
evaluated and gives a small contribution compared to that of the 
term $ln (1/|k_{x}|\ell_{0})$. The result is 
 
		\begin{equation}
		\{-(\omega-k_{x}v_{g})+S\ [ln 
		\frac{1}{|k_{x}\ell_{0}|}+\frac{3}{4}]\}\\*
		\ \int_{-\infty}^{\infty}dy\ \rho(\omega, k_{x},y)=0.
		\label{15}
		\end{equation}
From Eq. (\ref{15}) it follows that the EMP dispersion relation, with 
strong dissipation at the edges and for $k_{B}T\ll \hbar 
v_{g}/\ell_{0}$, is given by ($\omega (k_{x})\equiv \omega$) 
		
		\begin{equation}
		\omega = k_{x} v_{g}+\frac{2}{\epsilon}[k_{x}\sigma_{yx}^{0}
		-i k_{x}^{2}\tilde{\sigma}_{xx}]\ [ln \frac{1}{|k_{x}\ell_{0}|}+\frac{3}{4}].
		\label{16}
		\end{equation}
For $\nu=2$ the EMP dispersion relation will 
again be given by Eq. (\ref{16}) with the conductivity components 
multiplied by a factor of 2. In addition, because $v_{g}$ has the $\sim 1/B$ 
dependence, cf. Eq. (\ref{12}), it will be multiplied, for $\nu=2$,  
by a factor of 2 if the edge field $E_{e}$ is the same. As a result, for $\nu=2$ 
the frequency $\omega$ will be approximately twice larger than for 
$\nu=1$. More exactly, the value of the ratio of these frequencies is

		\begin{equation}
		\frac{\omega (\nu=2)}{\omega (\nu=1)}= 2 \{1-
		\frac{ln 2}{2[ln (1/|k_{x}\ell_{0}|)+3/4]}\},
		\label{17}
		\end{equation}
where $\ell_{0}$ corresponds to $\nu=1$. For $k_{x}\ell_{0}\sim 
10^{-6}$, the second  term inside the curly brackets represents a 
$2\%$ correction.

Eq. (\ref{16}) is valid for an ungated sample. If the sample is 
gated, repeating the procedure leads again to Eq. (\ref{16}) with the 
factor $[\ldots]$ replaced by $[ln (2d/ \ell_{0}|)+2/\pi]$. If the 
gate is replaced by air, this factor is replaced by $[ln (1/k_{x}^{2}d 
\ell_{0}|) ]$.

\section{NOVEL EDGE WAVES AT VERY LOW TEMPERATURES}

From Eq. (\ref{11}) it follows that, even for $T\rightarrow 0$,
$\rho(\omega,k_{x},y)=\rho^{(0)}(\omega,k_{x})\Psi_{0}^{2}(\bar{y})$, 
$\bar{y}=y-y_{re}$, 
is only an approximate solution of this equation. A more accurate solution is
obtained by the expansion
	\begin{eqnarray}
	\nonumber
		\rho(\omega,k_{x},y)=&&\Psi_{0}^{2}(\bar{y}) \sum_{n=0}^{\infty}
		\rho^{(n)}(\omega,k_{x}) H_{n}(\bar{y}/\ell_{0})\\*
		&&= \sum_{n=0}^{\infty}\sqrt{2^{n}n!}
		\rho^{(n)}(\omega,k_{x}) \Psi_{n}(\bar{y})\Psi_{0}(\bar{y}),
		\label{22}
		\end{eqnarray}
where $H_{n}(x)$ are the Hermite polynomials.  Due to their 
orthonormality Eq.  (\ref{22}) is the exact expression for 
$\rho(\omega,k_{x},y)$.  Notice that this expansion is specific to the 
case when only the lowest LL is occupied.  In addition, the terms 
$n=0, n=1, n=2$, etc.  correspond to the monopole, dipole, quadrupole, 
etc.  expansions of $\rho(\omega,k_{x},y)$ relative to $y=y_{re}$.

We now multiply Eq. (\ref{11}) by $H_{m}(\bar{y}/\ell_{0})$ 
and integrate over $y$ from  $y_{re}-\Delta y$ to $y_{re}+\Delta y$. 
Taking into account that for very low temperatures ( $\hbar 
v_{g}\gg \ell_{0}k_{B}T$) $d\sigma_{yx}^{0}(y)/d y$, $\sigma_{yy}(y)$, and 
$\sigma_{xx}(y)$ behave as $\Psi_{0}^2(\bar{y})$, we obtain
	\begin{equation}
		-(\omega-k_{x}v_{g})\rho^{(m)}(\omega, k_{x})+(S+mS')
		\sum_{n=0}^{\infty}(\frac{2^{n}n!}{2^{m}m!})^{1/2}a_{mn}( k_{x})\rho^{(n)}(\omega, k_{x})=0, 
		\label{23}
		\end{equation}
where

	\begin{equation}
	a_{mn}( k_{x})=a_{nm} ( k_{x})=\int_{-\infty}^{\infty} dx\ \Psi_{m}(x)\Psi_{0}(x)
	 \int_{-\infty}^{\infty} dx'\  K_{0}(|k_{x}||x-x'|)\ \Psi_{n}(x')\Psi_{0} (x')
	\label{24}
	\end{equation}
and $S'=-4i\tilde{\sigma}_{yy}/\epsilon\ell_{0}^{2}$. Notice that for $m=0$ Eq.  (\ref{23}) is equivalent to Eq.  (\ref{13}) 
and correspondingly the terms related to $\tilde{\sigma}_{yy}$ are 
absent. From Eqs.  (\ref{22}) -(\ref{24})  it follows that there exist 
independent wave modes, spatially {\it symmetric}, $\rho^{s}(\omega,k_{x},y)$, and 
{\it antisymmetric}, $\rho^{as}(\omega,k_{x},y)$, with respect to 
$y=y_{re}$.  They are given by  Eqs.  (\ref{22}) and  (\ref{23}) with 
$n$ {\it even} and {\it odd},
respectively.  Notice that in Eq.  (\ref{24}) due to the assumption $k_{x}\ell_{0}\ll 1$
we can write $K_{0}(|x|)\approx ln(2/|x|)-\gamma$ 
 for the ungated sample.  For the gated sample we 
simply replace the kernel $K_{0}$ in Eq. (\ref{24}) by $R_{g}\approx 
ln(2d/\ell_{0})-ln(|y-y^{`}|/\ell_{0})$ and for that with air by 
 $R_{a}\approx 
ln(2/k_{x}^{2}d\ell_{0}) -2\gamma -ln(|y-y^{`}|/\ell_{0})$. 

\subsection{Symmetric modes}

Considering only the term $n=0$ in the expression for 
$\rho^{s}(\omega,k_{x},y)$, Eq.  (\ref{23}) for $m=0$ gives
	
	\begin{equation}
	[-(\omega-k_{x}v_{g})+S a_{00} ]\rho^{(0)}(\omega, k_{x})=0, 
	\label{25}
	\end{equation}
where $a_{00}(k_{x})=-ln(|k_{x}|\ell_{0})+3/4$.  With this value of 
$a_{00}(k_{x})$ and $\rho^{(0)}(\omega, k_{x})\neq 0$, Eq.  (\ref{25}) 
gives the dispersion relation (\ref{16}).  Because $\rho^{s}(\omega, 
k_{x},y)$ behaves spatially as $\Psi_{0}^{2}(\bar{y})$ in this 
approximation, we will refer to it as the dispersion relation of the 
monopole EMP. For  the sample with a gate we simply have to replace 
$a_{00}$ by $a_{00}^{g}=[ln (2d/ \ell_{0})+2/\pi]$ in Eq.  
(\ref{25}) and for that with air by $a_{00}^{a}=ln (1/k_{x}^{2} \ell_{0}d)$.

Corrections to Eq. (\ref{25}) and further {\it symmetric} branches 
are obtained by keeping only the terms $n=0$ and $n=2$ in the 
expression for $\rho^{s}(\omega,k_{x},y)$ which gives
		\begin{equation}
		\rho^{s}(\omega,k_{x},y)=\rho^{(0)}(\omega,k_{x})\Psi_{0}^{2}(\bar{y})+
		2\sqrt{2}\rho^{(2)}(\omega,k_{x}) \Psi_{2}(\bar{y})\Psi_{0} (\bar{y}).
		\label{25a}
		\end{equation} 
From Eq.  (\ref{23})  for $m=0$ we obtain
		\begin{equation}
		[-(\omega-k_{x}v_{g})+S\ a_{00}]\rho^{(0)}(\omega, k_{x})+
		 2\sqrt{2}\ S\ a_{02}\rho^{(2)}(\omega, k_{x})=0, 
		\label{26}
		\end{equation}
and for $m=2$
	\begin{equation}
		[-(\omega-k_{x}v_{g})+(S+2S')
		\ a_{22}]\rho^{(2)}(\omega, k_{x})+
		[(S+2S')/2\sqrt{2}] a_{02} \rho^{(0)}(\omega, k_{x})=0, 
		\label{27}
		\end{equation}
where we write $a_{mn}(k_{x})\equiv a_{mn}$ in order to simplify the notation.  
 For a nontrivial solution of the system of Eqs.  (\ref{26}) and 
 (\ref{27}) the $2\times2$ determinant of the coefficients must 
 vanish.  This gives two branches $\omega_{+}^{s}(k_{x})$ and 
 $\omega_{-}^{s}(k_{x})$.  For $|k_{x}|\ell_{0}\ll 1$ a numerical 
 evaluation gives $a_{02} =-0.353$, $a_{22} =0.250$, and $a_{02}^{2} 
 =1/8$.  All $a_{mn}$ values remain the same for gated samples or 
 those with air except $a_{00}$ which changes as indicated 
 above.  If we neglect the coupling terms, by formally setting 
 $a_{02}(k_{x})=0$, Eq.  (\ref{26}) gives the monopole EMP dispersion 
 relation (\ref{16}) and Eq.  (\ref{27}) the pure quadrupole EMP 
 dispersion relation

		\begin{eqnarray}
		\nonumber
		\omega=&&k_{x}v_{g} + (S +2S')/4\\* 
		=&&k_{x}v_{g}+\frac{1}{2\epsilon}[k_{x}\sigma_{yx}^{0}-i k_{x}^{2}\tilde{\sigma}_{xx} 
		-4i\frac{\tilde{\sigma}_{yy}}{\ell_{0}^{2}}].
		\label{28}
		\end{eqnarray}
If we neglect $k_{x}v_{g}$ and the dissipative terms, Eq.  (\ref{28}) 
takes the form of Eq.  (14) of Ref.  \cite{11} for the $j=4$ 
branch which has $5$ charge oscillations.  As it stands Eq.  
(\ref{28}) corresponds to only $3$ oscillations.  The difference is to 
be ascribed to the very different density profile used in Ref.  
\cite{11} for a compressible liquid in a very wide strip.  Notice that 
Eq.  (\ref{28}) is valid for  samples with gate or air as well.

For finite $a_{02}$ the two branches resulting from Eqs. 
(\ref{26}) and  (\ref{27}) are given by 
	\begin{eqnarray}
	\nonumber
		\omega_{\pm}^{s}=&&k_{x}v_{g}+(1/2)[S(a_{00} +a_{22})+2S'a_{22}] \\*
		&&\pm (1/2)\sqrt{[S (a_{00}-a_{22}) -2S'a_{22}]^{2}
		+4S (S  +2S')a_{02}^{2}} . 
		\label{29}
		\end{eqnarray}
If we put $a_{02}=0$, i.e., if we neglect the coupling between the 
branches, then the $\omega_{-}^{s}(k_{x})$ branch is given by Eq.  
(\ref{28}) and the $\omega_{+}^{s}(k_{x})$ branch coincides with Eq.  
(\ref{16}).  It can be shown that the term $\propto a_{02}^{2}$ under 
the square root sign is much smaller than the other term.  Then from 
Eq.  (\ref{29}) we obtain

		\begin{equation}
		\omega_{+}^{s}=k_{x}v_{g}+S a_{00}
		+\frac{S (S + 2S')a_{02}^{2}}{Sa_{00}-(S + 
		2S')a_{22}}, 
		\label{30}
		\end{equation}  
and
\begin{equation}
		\omega_{-}^{s} =k_{x}v_{g}+(S+2S') a_{22}
		-\frac{S(S + 2S')a_{02}^{2}}{Sa_{00} -(S+ 
		2S')a_{22}}. 
		\label{31}
		\end{equation}

Further, for these very low temperatures we can distinguish between (i) 
strong dissipation, for which $k_{x} \sigma_{yx}^{0}\ll 
4\tilde{\sigma}_{yy}/\ell_{0}^{2}$ and (ii) weak dissipation, for which 
$k_{x}\sigma_{yx}^{0}\gg 4\tilde{\sigma}_{yy}/\ell_{0}^{2}$.  Notice 
that the damping of the purely quadrupole EMP-Eq.  (\ref{28})- is such 
that in case (i) we have $|\Im \omega |\gg |\Re\omega |$ whereas in 
case (i) the opposite inequality holds.  The damping of Eq.  
(\ref{28}) is determined by the dissipative conductivities 
$\sigma_{yy}(y)$ and $\sigma_{xx}(y)$.  The two contributions differ 
by a very small factor $k_{x}^{2}\ell_{0}^{2}$.  As a result, the 
damping of the wave, $\propto k_{x}^{2}\tilde{\sigma}_{xx}$, can be 
usually neglected. Notice that Eqs.  (\ref{26})-(\ref{31}) are valid
for gated or ``air'' samples as well  with $a_{00}$ replaced by $a_{00}^{g}$
or  $a_{00}^{a}$.


For definiteness in numerical estimates 
we will use parameters pertinent to  GaAs/AsAlGaAs 
heterostructures.  As will be shown below, both cases (i) and (ii) are 
experimentally realized depending on the values of $v_{g}$ and $T$. 
For $v_{g}\alt s$ the damping in Eq.  (\ref{28}) is exponentially 
suppressed. The condition  $v_{g}<s$ requires a smooth energy 
dispersion near the edges. This possibility exists in the Hartree 
approximation for the confining potential but not in the Hartree-Fock 
approximation where the exchange leads to a logarithmically divergent  
$v_{g}$ \cite{16}. However, when correlations are taken into account,
a smooth energy dispersion results near the edges and $v_{g}$ is 
small \cite{16}.
In GaAs-based heterostructures the most common case is $v_{g}>s=2.5\times 
10^{5}$cm/sec. In this case using  Eq.  (\ref{14}) for $\nu=1$, 
$c'=\hbar (eh_{14})^{2}/2\rho_{V}s, h_{14}=1.2\times 10^{7}$ V/cm, and
$\rho_{V}=5.31$ gm/cm$^{3}$, we obtain

		\begin{equation}
		\frac{\tilde{\sigma}_{yy} }{\ell_{0}\sigma_{yx}^{0}}
		=0.16 \tilde{T}^{3}\tilde{B}^{-3/2}\tilde{v}_{g}^{-4},		 
		\label{32}
		\end{equation}
where $\tilde{T}=T/1^{o}K, \tilde{B}=B/(10$ Tesla), and 
$\tilde{v}_{g}=v_{g}/s$.  Eq.  (\ref{32}) is valid for $\nu=2$ as well 
because of the scaled quantities.  For $\tilde{T}=1, \tilde{B}=1$, and 
$\tilde{v}_{g}=2$ Eq.  (\ref{32}) gives $ 
\tilde{\sigma}_{yy}/\ell_{0} \sigma_{yx}^{0}=10^{-2}$.  The 
estimated   \cite{17} field $E_{e}$ leads, for $\tilde{B}=1$,  to 
$v_{g} \approx 4\times 10^{5}$cm/sec $>s$.  Further, if we assume that 
$\tilde{\sigma}_{yy}/\ell_{0}$ gives approximately the value of the diagonal 
conductivity in the edge strip of width $\ell_{0}\approx 80$\AA, then, 
because for strong magnetic fields ($\omega_{c}\tau^{*}\gg 1$) 
$\tilde{\sigma}_{yy} /\ell_{0}\sigma_{yx}^{0}\approx 
1/\omega_{c}\tau^{*}=10^{-2}$, we obtain an effective scattering rate 
$1/\tau^{*}\approx \omega_{c}\tilde{\sigma}_{yy} /\ell_{0}\sigma_{yx}^{0}
\approx  2.6\times 10^{11}$/sec in this strip.  This is approximately ten times 
larger than the scattering rate for a mobility $\mu=10^{6}$ cm$^{2}$/V sec.

From Eq.  (\ref{28}) we obtain

	\begin{equation}
		\frac{|\Re \omega| }{|\Im \omega |}
		\approx
		\frac{\sigma_{yx}^{0}|k_{x}|\ell_{0}}{4\tilde{\sigma}_{yy}/\ell_{0}}
			\approx 1.5 \tilde{T}^{-3}\tilde{B}^{-3/2}\tilde{v}_{g}^{4}
			|k_{x}|\ell_{0},		 
		\label{33}
		\end{equation}
where we assumed again that the term $v_{g}k_{x}$ can be neglected.  
Then, for $\tilde{T}=1, \tilde{B}=1$, and $\tilde{v}_{g}=2$, the RHS 
of Eq.  (\ref{33}) is approximately equal to $25 |k_{x}|\ell_{0}$.  
Only for $1\gg |k_{x}|\ell_{0}>4\times 10^{-2}$ does the quadrupole 
EMP become weakly dissipative.  For $\tilde{B}=1$, lower temperature 
$\tilde{T}=0.1$, and steeper confinement $\tilde{v}_{g}=4$, Eq.  
(\ref{33}) gives $|\Re \omega | /|\Im \omega |\approx 5\times 
10^{5}|k_{x}|\ell_{0}$.  In this case the quadrupole wave is very 
weakly damped for $1.25 \times 10^{6}$cm$^{-1}\gg |k_{x}|>2.5$ 
cm$^{-1}$.  In this region the implicit low-frequency condition $|\omega |\ll 
\omega_{c}$ is well satisfied since it corresponds to $|k_{x}|\ll 
1.6\times 10^{7}$cm$^{-1}$.

\subsection{Edge helicons}

We now analyze further the general formulas of subsection A. We first 
assume that $\sigma_{yx}^{0}|k_{x}|K\gg \tilde{\sigma}_{yy}/
\ell_{0}^{2}$. Then  Eq.  (\ref{31}) gives

		\begin{equation}
		\omega_{-}^{s}=k_{x}v_{g} +\frac{1}{4}\ (S+2S')\ 
		(1-\frac{1}{2K}),
		\label{34}
		\end{equation}
where $K=a_{00}-1/4=ln(1/|k_{x}|\ell_{0})+1/2$; for gated or ``air'' 
samples $a_{00}$ in $K$ 
is replaced by $a_{00}^{g}$ and $a_{00}^{a}$, respectively.  Because 
$ln(1/|k_{x}|\ell_{0})\gg 1$, we see that the coupling with the 
monopole EMP does not change the dispersion almost at all as compared 
with that given by Eq.  (\ref{28}).  As a result the quasi-quadrupole 
EMP, described by Eq.  (\ref{34}), is weakly damped for 
$\sigma_{yx}^{0}|k_{x}|> 4\tilde{\sigma}_{yy}/ \ell_{0}^{2}$ and 
strongly damped for $4\tilde{\sigma}_{yy}/ 
\ell_{0}^{2}>\sigma_{yx}^{0}|k_{x}|\gg \tilde{\sigma}_{yy}/ 
\ell_{0}^{2}K $.  We call the wave described by Eq.  (\ref{34}) 
quasi-quadrupole EMP because it follows from Eq.  (\ref{27}) that

		\begin{equation}
		\frac{\rho^{(2)}(\omega, k_{x})}{\rho^{(0)}(\omega, k_{x})}=
		-2\sqrt{2}a_{02}K=K\gg 1; 
		\label{35}
		\end{equation}
that is, $\rho^{(2)}(\omega, k_{x})$ is  the dominant term on the RHS of 
Eq.  (\ref{25a}). The same holds for the ``air'' sample. However, for the 
gated sample the monopole and quadrupole terms are comparable if 
$K_{g}\leq 3$.
The condition of very weak damping for the wave (\ref{34}) can also be 
expressed as $\omega_{-}^{s}\tau^{*}\gg\nu r_{0}/\pi$ where $r_{0}=e^{2}/ 
\epsilon \hbar\omega_{c}\ell_{0}$. For $\nu=1$ we typically have
$r_{0}\sim 1$.  This  condition resembles the high-frequency 
limit used in Ref.  \cite{1}.  However, here $\tau^{*}$ is related to 
dissipation processes only near the edge.  In addition, in contrast 
with Ref.  \cite{1} we consider an essential  decrease of the conductivity 
components and of the electron density   over a finite length 
$\Lambda_{y}$ from the edge.  For very low temperatures, 
$k_{B}T\ll\hbar v_{g}/\ell_{0}$, we have $\Lambda_{y}\approx\ell_{0}$ 
which is much smaller than the length over which the density 
$n_{0}(y)$ decreases substantially in the model of Ref.  \cite{11}.

For $\sigma_{yx}^{0}|k_{x}|K\gg \tilde{\sigma}_{yy}/
\ell_{0}^{2}$ Eq.  (\ref{30}) gives  
		\begin{equation}
		\omega_{+}^{s}=k_{x}v_{g} +S\ 
		(K+1/4)[1+\frac{1}{8K(K+1/4)}]+\frac{S'}{4K}. 
		\label{36}
		\end{equation}
As can be seen, taking into account the coupling with the quadrupole 
EMP changes the phase velocity of the monopole EMP by a very small 
amount ($\leq 0.1\%$) but it gives a principally new contribution to
the damping in comparison with Eq.  (\ref{16}).  Now the $\omega_{+}^{s}$ 
branch, for the typical EMP condition $|k_{x}|\ell_{0}\sim 10^{-6}$, 
has a damping $\propto\tilde{\sigma}_{yy}/ \ell_{0}^{2}K$ which is much 
stronger than that $\propto\tilde{\sigma}_{xx} k_{x}^{2}K$ of the pure 
monopole EMP. 

The wave described by Eq.  (\ref{36}) can be called a quasi-monopole EMP 
because it follows from Eq.  (\ref{26}) that

		\begin{equation}
		\frac{\rho^{(0)}(\omega, k_{x})}{\rho^{(2)}(\omega, k_{x})}=
		\frac{16\sqrt{2}a_{02}\sigma_{yx}^{0}k_{x}
		K}{\sigma_{yx}^{0}k_{x}-4i\tilde{\sigma}_{yy}/\ell_{0}^{2}},
		\label{37}
		\end{equation}
and, due to $K\gg 1$, we 
have $|\rho^{(0)}(\omega, k_{x})/\rho^{(2)}(\omega, k_{x})|\gg 1$.  Now, 
 for weak dissipation ($\sigma_{yx}^{0}k_{x}\gg 
4\tilde{\sigma}_{yy}/\ell_{0}^{2}$) we have $\rho^{(0)}(\omega, 
k_{x})/\rho^{(2)}(\omega, k_{x})\approx - 8 K$ and  for strong 
dissipation ($\sigma_{yx}^{0}k_{x}K\gg 
\tilde{\sigma}_{yy}/\ell_{0}^{2}\gg \sigma_{yx}^{0}k_{x}/4$) 
$\rho^{(0)}(\omega, k_{x})/\rho^{(2)}(\omega, k_{x})\approx -2\ i 
(\sigma_{yx}^{0}k_{x} \ell_{0}^{2}/\tilde{\sigma}_{yy})K$.  Thus, if 
the phases of the two components are shifted by $\pi$ in the first 
case, in which we call the $\omega_{+}^{s}$ branch described by Eq.
(\ref{36}) a modified monopole  EMP (MMEMP),
they are shifted by $\pi/2$ in the last one.  This last case 
corresponds to the frequency region $\omega_{+}^{s}\tau^{*}\gg \nu 
r_{0}/\pi\gg (\omega_{+}^{s}\tau^{*}/(4K+1))$ and the frequency $\omega_{+}^{s}$
 can still be considered as high compared to $1/\tau^{*}$.  In this 
 frequency region we call the $\omega_{+}^{s}$ branch
 the high-frequency {\it edge helicon} (HFEH) and denote it by  
 $\omega_{EH}^{HF}$. In this 
 region, due to the almost $\pi/2$ shift 
between $\rho^{(0)}$ and $\rho^{(2)}$, we obtain the following 
remarkable property of HFEH described by Eq. (\ref{36}). 
If the HFEH charge $\propto Re\{\rho^{s}(\omega_{EH}^{HF}, 
k_{x}, y) R_{+}(x, t)\}$ along  $y$ has a pure quadrupole character $\propto 
|\rho^{(2)}|$ for some phase of the
 running wave $R_{+}(x, t) =  \exp[-i(\omega_{EH}^{HF} t-k_{x} x)]$, then 
 after approximately a $\pm\pi/2$ shift it acquires  
a  pure monopole character $\propto |\rho^{(0)}|$;
$\rho^{s}(\omega_{EH}^{HF}, 
k_{x}, y)$ is given by Eq. (\ref{25a}). This HFEH shows {\it three} 
charge oscillations along $y$ whereas 
the
relevant branches of Refs. \cite{1} and 
\cite{11}, with $Re \omega\propto k_{x}ln (1/k_{x})$ resembling
$\Re \omega^{HF}_{EH}$ of the HFEH,   show only {\it one} oscillation. 

Notice that $\Re \omega_{+}^{s}$, given by Eq.  (\ref{36}) is independent 
of $T$ whereas $\Im \omega_{+}^{s}\propto T^{3}$ or $T^{2}$ if $1\gg 
k_{B}T \ell_{0}/\hbar v_{g}\gg s/\sqrt{2}v_{g}$.  That is, in contrast with 
Ref.  \cite{1} we find that the damping of the MMEMP and 
 that of the HFEH 
scale  with temperature and are not quantized in the QHE plateaus.  In 
addition, these waves have a characteristic length $\ell_{0}$ which is 
different than the length $\ell_{v}$ of Ref. \cite{1}; 
 also,  the term $\propto \sigma_{yx}^{0}$ is different than the that
of Ref. \cite{1} in the  factor containing the logarithm.
In addition, $\ell_{0}(\nu=2)/\ell_{0}(\nu=1)=\sqrt{2}$ 
here whereas $\ell_{v}(\nu=2)/\ell_{v}(\nu=1)=4$ in the high frequency
limit of Ref.  \cite{1}.  
Moreover, for 
$\epsilon\rightarrow \infty $ it follows that $\ell_{v}(\nu=1)\rightarrow 0$
 whereas $\ell_{0}$ is independent of $\epsilon$.

We now consider the case of very strong dissipation
		\begin{equation}
		\tilde{\sigma}_{yy}/\ell_{0}^{2}\gg \sigma_{yx}^{0}|k_{x}|K.
		\label{38}
		\end{equation}
Then Eq.  (\ref{31}) gives
	\begin{equation}
		\omega_{-}^{s}=k_{x}v_{g} +\frac{1}{2\epsilon}
		\ [3\sigma_{yx}^{0}k_{x}-4i\tilde{\sigma}_{yy}/\ell_{0}^{2}].
		\label{39}
		\end{equation}  
This is again a quasi-quadrupole wave since
$|\rho^{(2)}(\omega, k_{x})/\rho^{(0)}(\omega, k_{x})|\approx 
  \tilde{\sigma}_{yy}/\sigma_{yx}^{0}|k_{x}| \ell_{0}^{2}\gg 1$. 
Also, although  $\Re \omega_{-}^{s}/k_{x}$ is essentially larger than in 
Eq.  (\ref{34}), Eq.  (\ref{38}) gives an {\it aperiodic damping},
$|\Re \omega_{-}^{s}|\ll |\Im \omega_{-}^{s}|$. Further, assuming that   Eq.  (\ref{38}) is 
valid, we obtain from  Eq.  (\ref{30}) the dispersion relation of a 
new wave that we call low-frequency {\it edge helicon} (LFEH) as
	\begin{equation}
		\omega_{EH}^{LF}=k_{x}v_{g} +S(K-\frac{1}{4}) 
		-\frac{i}{\epsilon}\frac{[\sigma_{yx}^{0}k_{x}]^{2}}
		{\tilde{\sigma}_{yy}/\ell_{0}^{2}}(K-\frac{1}{4}).
		\label{40}
		\end{equation}
Despite the strong dissipation condition (\ref{38}), which entails 
$\omega^{LF}_{EH}\tau^{*}\ll \nu r_{0}/\pi\alt 1$, the LFEH is very 
weakly damped since $|\Re \omega_{EH}^{LF}|\gg |\Im \omega_{EH}^{LF}|$.  The 
frequency range of the LFEH is similar to the low-frequency limit of 
Ref.  \cite{1} but here $\tau^{*}$ is related to strong dissipation 
processes only near the edges.  Also, $\Re \omega_{EH}^{LF}/k_{x}$ 
differs little from $\Re \omega_{EH}^{HF}/k_{x}$ 
 or that of Eq.  (\ref{16}).  However, the 
damping of the LFEH $\propto 
(\sigma_{yx}^{0}k_{x}\ell_{0})^{2}ln(1/k_{x}\ell_{0})/\tilde{\sigma}_{yy}$ 
has a very different form than $\Im \omega_{EH}^{HF}$ of Eq.  
(\ref{36}) or $\Im \omega$ of Eq.  (\ref{16}).

We further notice that, in contrast with Ref.  \cite{1}, the real part 
$\Re \omega_{EH}^{LF}$ is independent of temperature whereas the 
imaginary part $\Im \omega_{EH}^{LF}$, i.  e., the damping, is not 
quantized and varies as $T^{-3}$ or $T^{-2}$; the $T^{-2}$ behavior
occurs if $v_{g}^{2}\gg s^{2}$ and $1\gg k_{B}T 
\ell_{0}/\hbar v_{g}\gg s/\sqrt{2}v_{g}$.  That is, the LFEH has a 
characteristic length very different than $\ell_{v}$ in the 
low-frequency limit.  It follows from Eq.  (\ref{26})  that

	\begin{equation}
		\frac{\rho^{(0)}(\omega_{EH}^{LF}, k_{x})}{\rho^{(2)}(\omega_{EH}^{LF},
		 k_{x})}
		\approx -4\sqrt{2}a_{02}= 2.
		\label{41}
		\end{equation}
With this result and Eq.  (\ref{25a}) we obtain the dimensionless 
charge density profile of the LFEH, $ 	
\tilde{\rho}_{EH}(y)=\sqrt{\pi}\ell_{0}
\rho^{s}(\omega_{EH}^{LF}, k_{x}, y)/\rho^{(0)}(\omega_{EH}^{LF}, k_{x})$, as
		\begin{equation}
		\tilde{\rho}_{EH}(y)=\sqrt{\pi}\ell_{0}
		[\Psi_{0}^{2}(\bar{y}) +\sqrt{2}\Psi_{2} (\bar{y})
		\Psi_{0}(\bar{y})].
		\label{42}
		\end{equation}  

In Fig.  2 we show $ \tilde{\rho}_{EH}(y)$ (solid curve ), its 
monopole component (term $ \propto \Psi_{0}^{2}$, short-dashed curve), 
and its quadrupole component (term $\propto \Psi_{0} \Psi_{2}$, 
long-dashed curve ).  For contrast the dotted curve represents the 
normalized 
unperturbed electron density $n_{0}(y)/n_{0}$.  As can be seen the 
monopole and quadrupole contributions are of the same order of 
magnitude and the resultant $ \tilde{\rho}_{EH}(y)$ has an oscillatory 
behavior with two oscillations, one to the right and one to the left of the edge
at $y=y_{re}$.  
This is in sharp contrast with the ``usual'' EMPs of Refs.  \cite{1} 
and the $j=0$ mode of Ref.  \cite{11}.

\subsection{Antisymmetric modes}

Considering only the term $n=1$ in the expression for $\rho^{as}(\omega, 
k_{x}, y)$ and using  Eq.  (\ref{23})  for $m=1$ we obtain

	\begin{equation}
	[-(\omega-k_{x}v_{g})+(S+S') a_{11}]\rho^{(1)}(\omega, k_{x})=0; 
	\label{43}
	\end{equation}
the numerically obtained value of $a_{11}$ is $0.5$. For $\rho^{(1)}(\omega, k_{x})
\neq 0$ Eq.  (\ref{43})  gives the dispersion relation of the pure
dipole EMP as

	\begin{equation}
	\omega=k_{x}v_{g}+ (1/\epsilon)[k_{x}\sigma_{yx}^{0} 
	-2i\tilde{\sigma}_{yy}/\ell_{0}^{2}]. 
	\label{44}
	\end{equation}
If we neglect $k_{x}v_{g}$ and the dissipative term,   Eq.  (\ref{44})  
will take
the form of Eq. (14) of Ref. \cite{11} for the $j=2$ which shows three charge 
oscillations whereas Eq.  (\ref{44}) corresponds to only two
oscillations. However, in contrast with Ref. \cite{11}, besides the 
term $k_{x}v_{g}$ and the microscopically treated dissipative term, 
the Hall conductivity $\sigma_{yx}^{0}$ is quantized for $\nu=1$ or 
$2$. 

Corrections to Eqs. (\ref{43})  and  (\ref{44}) and further
antisymmetric branches are obtained by keeping only the $n=1$ and 
$n=3$ terms in the expression for  $\rho^{as}(\omega, 
k_{x}, y)$. Then

	\begin{equation}
		\rho^{as}(\omega,k_{x},y)=
		\sqrt{2}\rho^{(1)}(\omega,k_{x}) \Psi_{1}(\bar{y})\Psi_{0} (\bar{y})
		+4\sqrt{3}\rho^{(3)}(\omega,k_{x}) \Psi_{3}(\bar{y})\Psi_{0} (\bar{y}).
		\label{45}
		\end{equation}
From Eqs. (\ref{23})  and (\ref{24}), for $m=1$, we obtain

	\begin{equation}
		[-(\omega-k_{x}v_{g})+(S+S')\ a_{11}]\rho^{(1)}(\omega, k_{x})+
		2\sqrt{6}(S+S')\ a_{13}\rho^{(3)}(\omega, k_{x})=0, 
		\label{46}
		\end{equation}
and for $m=3$
	\begin{equation}
		[-(\omega-k_{x}v_{g})+(S+3S')
		\ a_{33}]\rho^{(3)}(\omega, k_{x})+
		(1/2\sqrt{6})(S+3S') a_{13} \rho^{(1)}(\omega, k_{x})=0. 
		\label{47}
		\end{equation}
Again the vanishing of the $2\times 2$ determinant of the coefficients gives
the two branches $\omega_{-}^{as}$ and $\omega_{+}^{as}$. In the long 
wavelength limit we numerically evaluate $a_{13}=-0.204$ and 
$a_{33}=0.166=1/6$. If we neglect the coupling between the modes, by 
formally setting $a_{13}=0$, Eq. (\ref{46})  gives the dispersion relation 
for the pure dipole EMP,  Eq. (\ref{44}), and  Eq. (\ref{47}) that for the octupole EMP

\begin{equation}
	\omega=k_{x}v_{g}+(1/3\epsilon)
	[k_{x}\sigma_{yx}^{0} -6i\tilde{\sigma}_{yy}/\ell_{0}^{2}]. 
	\label{48}
		\end{equation}
If we neglect $k_{x}v_{g}$ and the dissipative term Eq. (\ref{48}) 
takes the form of Eq. (14) of Ref. \cite{11} for the $j=6$ branch
that shows seven charge oscillations. As
it stands, Eq. (\ref{48}) corresponds only to four charge oscillations.

For finite $a_{13}$ the two branches resulting from Eqs. (\ref{46})
and  (\ref{47}) are given by

\begin{eqnarray}
	\nonumber
	\omega_{\pm}^{as}=&&k_{x}v_{g}+(1/2)[S(a_{11} +a_{33})+S'(a_{11}+3a_{33})] \\*
		&&\pm (1/2)\sqrt{[S (a_{11}-a_{33}) +S'(a_{11}-3a_{33})]^{2}
		+4a_{13}^{2}(S  +S')(S  +3S')}. 
		\label{49}
		\end{eqnarray}
If we set  $a_{13}=0$, $\omega_{+}^{as}$ and $\omega_{-}^{as}$ give the dipole and octupole branches given 
above by Eqs. (\ref{44}) and (\ref{48}), respectively. For weak 
dissipation we have   
$k_{x}\sigma_{yx}^{0}\gg 4\tilde{\sigma}_{yy}/\ell_{0}^{2}$ and, if we 
neglect damping, we obtain

\begin{equation}
	\omega_{+}^{as}\approx k_{x}v_{g}+(6/5\epsilon)\sigma_{yx}^{0}k_{x} 
		\label{50}
		\end{equation}	
and
	\begin{equation}	
			\omega_{-}^{as}\approx k_{x}v_{g}+(1/7\epsilon)\sigma_{yx}^{0}k_{x}. 
		\label{51}		
		\end{equation}
We call the waves corresponding to Eqs. (\ref{50}) and (\ref{51}) 
modified dipole (MDEMP) and octupole (MOEMP) EMP, respectively. As we 
now show, some of their properties are essentially different than 
those of the   pure dipole (Eq. (\ref{44}))
and octupole (Eq. (\ref{48})) EMPs.

Then from Eq. (\ref{46}) for the MDEMP we calculate

	\begin{equation}	
			\frac{ \rho^{(1)}(\omega, k_{x})}{ \rho^{(3)}(\omega, 
			k_{x})}=20\sqrt{6}a_{13}
			=-10.0.
		\label{52}		
		\end{equation}
With this ratio and Eq. (\ref{45}) the dimensionless charge 
density profile  $ \tilde{\rho}_{\pm}^{as}(\omega_{\pm}^{as}, k_{x}, y)=
\sqrt{\pi}\ell_{0}\rho^{as}(\omega_{\pm}^{as}, k_{x}, 
y)/\rho^{(1)}(\omega_{\pm}^{as}, k_{x}, y)$ of the MDEMP takes the form

		\begin{equation}
	\tilde{\rho}_{MD}(y)\equiv\tilde{\rho}_{+}^{as}(y)=\sqrt{2\pi}\ell_{0}\ 
		[\Psi_{1} (\bar{y})\Psi_{0} (\bar{y}) -(1/2)\Psi_{3} (\bar{y})
		\Psi_{0}(\bar{y})].
		\label{53}
		\end{equation}  
In Fig. 3 we show  $\tilde{\rho}_{MD}(y)$ (solid curve), 
its dipole component (term $\propto\Psi_{1} (\bar{y})\Psi_{0}(\bar{y})$, short-dashed 
curve), and its octupole component (term $\propto\Psi_{3} (\bar{y})\Psi_{0}(\bar{y})$
long-dashed curve). Again the dotted curve represents the normalized unperturbed 
electron density $n_{0}(y)/n_{0}$. As can be seen the dipole and 
octupole contributions to $\tilde{\rho}_{MD}(y)$ are of the same 
order of
magnitude though $\omega_{+}^{as}\equiv\omega_{MD}$, given by Eq. 
(\ref{50}), is a bit 
different ($<20\%$) than $\Re \omega$ of the pure dipole EMP 
given by Eq. (\ref{44}). As seen in Fig. 3, the MDEMP has four charge
oscillations whereas the dipole EMP has two. Thus, the corresponding 
density 
profiles are qualitatively different although the phase velocities are close to each 
other.

For the MOEMP, given by Eq. (\ref{51}),  we calculate from Eq. 
(\ref{47})
	\begin{equation}	
			\frac{ \rho^{(3)}(\omega, k_{x})}{ \rho^{(1)}(\omega, k_{x})}
			=2.109/2\sqrt{6}\approx 1/\sqrt{6};
		\label{54}		
		\end{equation}	
the corresponding result for  $\tilde{\rho}_{MO} (y)\equiv\tilde{\rho}_{-}^{as} (y)$ is

	\begin{equation}
		\tilde{\rho}_{MO}(y)=\sqrt{2\pi}\ell_{0}\ 
		[\Psi_{1} (\bar{y})\Psi_{0} (\bar{y}) +2\Psi_{3} (\bar{y})
		\Psi_{0}(\bar{y})].
		\label{55}
		\end{equation} 
In Fig.  4 we plot the same quantities as in Fig.  3 but now for 
$\tilde{\rho}_{MO}(y)$.  As can be seen, the spatial behavior of 
$\tilde{\rho}_{MO}(y)$ is quantitatively different only than that of the pure 
octupole EMP. The phase velocities of these two EMPs, as follow from 
the dispersion relations, are substantially different.  Notice that 
for strong dissipation, $k_{x}\sigma_{yx}^{0}\ll 4 
\tilde{\sigma}_{yy}/\ell_{0}^{2}$, both  $\omega_{\pm}^{as}$ branches
are strongly damped. 
%
\section{DISCUSSION AND CONCLUDING REMARKS}

We have introduced a realistic model for the confining potential $V_{c}(y)$ and 
treated mainly the case where $\nu=1$ in the interior part of the 
channel and the formation of dipolar strips \cite{18} near the  edges is 
impossible in the assumed QHE regime. We have taken $V_{c}(y)$ 
sufficiently steep at the edge that LL flattening can be neglected
\cite{18}-\cite{20}. As for $\nu=2$, we have neglected the spin splitting. This 
is a reasonable approximation in the bulk of the channel but its 
validity near the edges is not clear in view of the work of Refs. 
\cite{16} and \cite{21}.  Though we have used a simple form for $V_{c}(y)$,
 near the edge the results remain  valid for potentials 
$V_{c}(y)$ of different form that are smooth on the scale of $\ell_{0}$. In such a 
case, e.g., the last term on the RHS of Eq. (\ref{5}) should be 
replaced by $V_{c}(y_{0\alpha})$ and $v_{g}$ will be given again by 
Eq. (\ref{12}) with $k_{re}$ determined by $f_{0}(k_{re})=1/2$. Thus, $v_{g}$
is the only parameter related to the form of $V_{c}$ that influences 
these edge waves.

For a comparison of the  theory with the experiment the details of the
sample geometry are necessary but often they are not given. For 
instance, the thickness $d_{s}$ of the sample is not given in Ref. 
\cite{6}; it is also not given neither in Ref. \cite{22},where the bottom 
of the sample is metallized, nor in Ref.  \cite{9}. Now concerning Ref.  \cite{5}
it can be seen that $d_{s}=400\ \mu$m for the sample with perimeter
$P=1.7\ $cm and the observations were made at $T=0.4$K and $B=13$ 
Tesla ($\nu=1$). As we now show, the first three edge helicon modes $k_{x0}=2\pi/P$,  $k_{x1}=2 k_{x0}$, and
$k_{x2}=3 k_{x0}$ do not lead to 
an equidistant spectrum. Indeed, in addition to the case of the sample with air
treated above, there is air for $z<-d_{s}$ (a foam plastic base with 
small dielectric constant). Then for  $2 k_{x}d_{s}\ll 1$ we arrive at 
Eq. (\ref{10}) with $\epsilon\approx 1$, cf. Ref. \cite{23}, which should be substituted in 
all formulas involving a homogeneous $\epsilon $; for  $2 
k_{x}d_{s}\gg 1$ we obtain the case of the ``air'' sample treated above. Taking into account 
that $2k_{x0}d_{s}=4\pi d_{s}/P\sim 0.3$, $\ 2k_{x1}d_{s}\sim 0.6$, and
$2k_{x2}d_{s}\sim 0.9$, we can explain qualitatively the observed, not 
equidistant spectrum. A quantitative comparison is impossible because 
the condition $|k_{x}|W\gg 1$ is not satisfied ($|k_{x}|W\sim 1$). 

The  sample of Ref. \cite{8} is a circular mesa
 with diameter $D=540\ \mu$m, height $\sim 1 \mu$m, in the middle of
 a wide GaAs/AlGaAs chip with thickness $d_{s}=500\ \mu$m. The
condition $2 |k_{x}|d_{s}\gg 1$ is well satisfied even for 
$k_{x0}=2/D=37$ cm$^{-1}$; the minimum value of $k_{x}$, $k_{xmin}$, 
involved in the experiment satisfies $k_{xmin}\gg k_{x0}$. As a result, 
the conditions $2 k_{xmin}d_{s}\gg 1$ and $k_{xmin} W\gg 1$ are well satisfied as well. Because the
square ``pulser'' gate, of width $L_{p}=10\mu$m was much smaller than 
the circumference $\pi D$, the initial charge distribution can be 
assumed to have a rectangular form and therefore to have an essential 
contribution  from the $k_{xn}=\pm n k_{x0},\ n=1,2,..,$ modes 
distributed in the interval $k_{x0}\leq |k_{xn}|<\pi/L_{p}$.
It is natural to assume that the typical $k_{xt}=|k_{xnt}|\approx 
\pi/2L_{p}$; 
for a rectangular form of the   charge distortion along $x$, we have 
approximately a $50\%$ contribution to the total spectral density for
$|k_{xn}|\leq k_{xt}$. The 
model of the ``air'' sample, described briefly in the Appendix, fits 
perfectly to the experiment  \cite{8}. The distance $d$ at which the 
2DEG is situated beneath the surface is not given but from the mesa 
height it can be 
estimated to be $d\approx1000\AA$. This gives $K_{a}(k_{xt})\approx 
10.5$ for $B=5.1$ Tesla. Further, assuming $\tilde{v}_{g}=2$, i.e., $v_{g}=5\times 
10^{5}$ cm/sec, using $\tilde{B}=0.51, \ \tilde{T}=0.3$, and 
$\nu=1$, in the middle of a wide chip, obtained from  the  parameters of the experiment, and using 
Eq. (\ref{32}), we obtain $\tilde{\sigma}_{yy}/\ell_{0}\sigma_{yx}^{0}
\approx 7.5\times 10^{-4}$. 
It follows that  $ k_{xt}\ell_{0}K_{a}(k_{xt})\approx 1.6\times 10^{-2}\gg
\tilde{\sigma}_{yy}/\ell_{0}\sigma_{yx}^{0}$ which corresponds to the 
first condition treated in Sec. IV. B. 

Notice that here $ k_{xt}\ell_{0}\approx 1.5\times 10^{-3}<
4\tilde{\sigma}_{yy}/\ell_{0}\sigma_{yx}^{0}\approx 3\times 10^{-3}$, i.e., 
we are dealing with the case of strong dissipation as detailed after 
Eq. (\ref{31}). It follows that all modes, 
apart from the HFEH, are strongly damped. For instance, both the
pure dipole (Eq. (\ref{44})) and quasi- quadrupole (Eq. (\ref{34})) modes have $\Im \omega\approx 
3.4\times 10^{10}$/sec which is a very large value compared to the
damping rate of the experiment $\Im \omega\approx 
2\times 10^{8}$/sec for $\nu=1$. Moreover, during the observed period for the travel 
around the edge  $T_{e}\approx 3.5\times 10^{-9}$ sec the amplitude 
of the quasi- quadrupole EMP or dipole EMP should practically vanish 
due to the exponentially small factor $<exp(-100)$. We are left with the 
high-frequency edge helicons described by Eqs. (\ref{36}) and (\ref{37}).
Eq. (\ref{36}) gives a decay rate $\Im \omega_{EH}^{HF}\approx 
1.9\times 10^{8}$/sec which is in good agreement with the 
observations \cite{8}. The corresponding group velocity for the HFEH, 
obtained from Eq. (\ref{36}), is $v_{g}(k_{xt})\approx 
(2/\epsilon)\sigma_{yx}^{0}[ln(1/k_{xt}^{2}\ell_{0}d)-2]$ and gives a 
period $T=\pi D/v_{g}(k_{xt})\approx 3.4\times 10^{-9}$/sec which is
in excellent agreement with the experimental value. Because after the 
first trip the pulse became 700 ps wider, we can estimate the range of 
$\Delta k_{x}$, around $k_{xt}$, which gives the most essential 
contribution to the pulse. Subtracting  $v_{g}(k_{xt}+\Delta k_{x}/2)$
from $v_{g}(k_{xt}-\Delta k_{x}/2)$ leads, after some calculations, to 
$\Delta k_{x}/2k_{xt}\approx 
0.4$. This then gives approximately $  k_{xmax}=1.4  k_{xt}$ and
$  k_{xmin}=0.6  k_{xt}$.

Another important ingredient of our theory are the calculated
damping rates. As shown above, they agree well with the observed rates
\cite{8}. The effective $\tau^{*}$ for the work of Ref. \cite{8}
 is approximately  $2.5\times 10^{-11}$ sec. 
This is several decades shorter than that extracted from QHE measurements
which is of the order of $10^{-3}$ sec \cite{24}. The difference is to
be ascribed to the  fact that in our model the dissipation is 
localized near the edges whereas in that 
of Ref. \cite{1} the dissipation occurs throughout the channel 
homogeneously. The latter  is a reasonable assumption for relatively high 
temperatures which have not been considered in the present work.

Finally, it is worth noticing that in our microscopically
calculated dispersion relations the quantized Hall conductivity $\sigma_{yx}$ appears 
explicitly. Though not  shown here graphically, this accounts for the existence of  plateaus in the transit times of the 
signals, as a function of magnetic field, observed 
recently in Ref. \cite{22} and  accounted for by the replacement of the quantity by $\sigma_{yx}$. The 
present theory holds when only  n=0 LL is occupied. The coupling 
between edge excitations of different LLs will certainly affect the EMP modes
 presented here and
an appropriate extension of the theory is being planned. 

\acknowledgements

This work was supported by NSERC Grant No. OGP0121756. In addition, O 
G B acknowledges partial support by the Ukrainian SFFI Grant No. 
2.4/665.
\ \\
\appendix{APPENDIX}

As a model of a gated sample, we consider a 2DEG at $z=0$ 
with a metallic gate placed at a distance $z=d$ away from it and
with a  dielectric constant $\epsilon$  for  $ z< d$.
Taking the Fourier transform, for $z<d$, with respect to $x, y$, and $t$, of  
Poisson's equation for the time-dependent  charge density $\rho(x,y,z,t)$ gives

	\begin{equation}
		\epsilon[k^{2}-\frac{\partial^{2}}{\partial z^{2}}]
		\phi(\omega, k_{x}, k_{y}, z)=4\pi\rho(\omega, k_{x}, k_{y})\delta (z),
		\label{A1}
		\end{equation}
where $k^{2}=k_{x}^{2}+k_{y}^{2}$. For $z\leq 0$ we have 

\begin{equation}
		\phi_{-}(\omega, k_{x}, k_{y}, z)=A (\omega, k_{x}, k_{y})\ e^{kz},
		\label{A2}
		\end{equation}
and for $0\leq z< d$
		
		\begin{equation}
		\phi_{+}(\omega, k_{x}, k_{y}, z)=B (\omega, k_{x}, k_{y})e^{kz}+
		C (\omega, k_{x}, k_{y})\ e^{-kz}.
		\label{A3}
		\end{equation}
Two boundary conditions are $\phi_{+}(\omega, k_{x}, k_{y}, d)=0$ and
$\phi_{+}(\omega, k_{x}, k_{y}, +0)=\phi_{-}(\omega, k_{x}, k_{y}, -0)$. 
Integrating Eq. (\ref{A1}) from $z=-0$ to $z=+0$ gives the third condition

	\begin{equation}
		\epsilon[\frac{\partial \phi_{+}(\omega, k_{x}, k_{y}, z)}
		{\partial z }|_{z=+0}\ 
		-\ \frac{\partial \phi_{-}(\omega, k_{x}, k_{y}, z)}
		{\partial z}|_{z=-0}]
		=-4\pi\rho(\omega, k_{x}, k_{y}).
		\label{A4}
		\end{equation}
From Eqs. (\ref{A2})-(\ref{A4}) and the first two conditions  we obtain

	\begin{equation}
		\phi(\omega, k_{x}, k_{y}, z=0)
		=2\pi\rho(\omega, k_{x}, k_{y}) (1-e^{-2kd})/\epsilon k
		\label{A5}
		\end{equation}
which gives

	\begin{eqnarray}
	\nonumber
		\phi(\omega, k_{x}, y)=&&\frac{1}{2\pi}
		\int_{-\infty}^{\infty}\ e^{ik_{y}y}\phi(\omega, k_{x}, k_{y}, 
		z=0) dk_{y}\\*
		&&=\frac{2}{\epsilon}
		\int_{-\infty}^{\infty}[K_{0}(|k_{x}||y-y'|) -  K_{0}(|k_{x}|
		\sqrt{(y-y')^{2}+4d^{2}})] \rho(\omega, k_{x},y')\ dy^{`}. 
		\label{A6}
		\end{eqnarray}
Notice that for $d\rightarrow\infty$ Eq. (\ref{A6}) coincides with Eq. 
(\ref{10}).

If the gate is replaced by air, for $z> d$, a similar calculation 
leads again to Eq. (\ref{A6}) with the kernel replaced by 
$R_{a}=K_{0}(|k_{x}||y-y'|) + [(\epsilon-1)/(\epsilon +1)] K_{0}(|k_{x}|
		\sqrt{(y-y')^{2}+4d^{2}})$.

\begin{figure}
\caption{Unperturbed electron density $n_{0}(y)$, normalized to the bulk 
value $n_{0}$, as a function of $\bar{y}/\ell_{0}$ measured from the right 
edge taken as the origin. The solid and long-dashed
curves are obtained from the models of Refs. \protect\cite{1} and \protect\cite{11}, 
respectively, as explained in the text. 
The short-dashed curve is the profile of the present work.}  
\label{fig.1}
\ \\
\caption{Dimensionless charge density profile $\tilde{\rho}_{EH}(y)$ of the 
low-frequency edge helicon (LFEH) described
by Eq. (\protect\ref{42}) as a function of $\bar{y}/\ell_{0}$ (solid 
curve). The dotted curve represents $n_{0}(y)/ n_{0}$. The dashed and long-dashed curves
 are the pure monopole 
and quadrupole contributions, respectively. The oscillatory behavior of 
$\tilde\rho_{EH}(y)$
near the edge is in 
sharp contrast with the ``usual'' EMP of Refs.  \protect\cite{1} and 
\protect\cite{11}.}
\label{fig.2}
\ \\
\caption{Dimensionless charge density profile $\tilde{\rho}_{MD}(y)$ 
 of the MDEMP described
by Eq. (\protect\ref{53}) as a function of $\bar{y}/\ell_{0}$ (solid 
curve). The dashed and 
long-dashed curves are the pure
dipole 
and octupole contributions, respectively.
 The dotted curve represents $n_{0}(y)/ n_{0}$.} 
\label{fig.3}
\end{figure}
\begin{figure}
\caption{As in Fig. 3 but for $\tilde{\rho}_{MO}(y)$ as described
by Eq. (\protect\ref{55}).} 
\label{fig.4}
\end{figure}
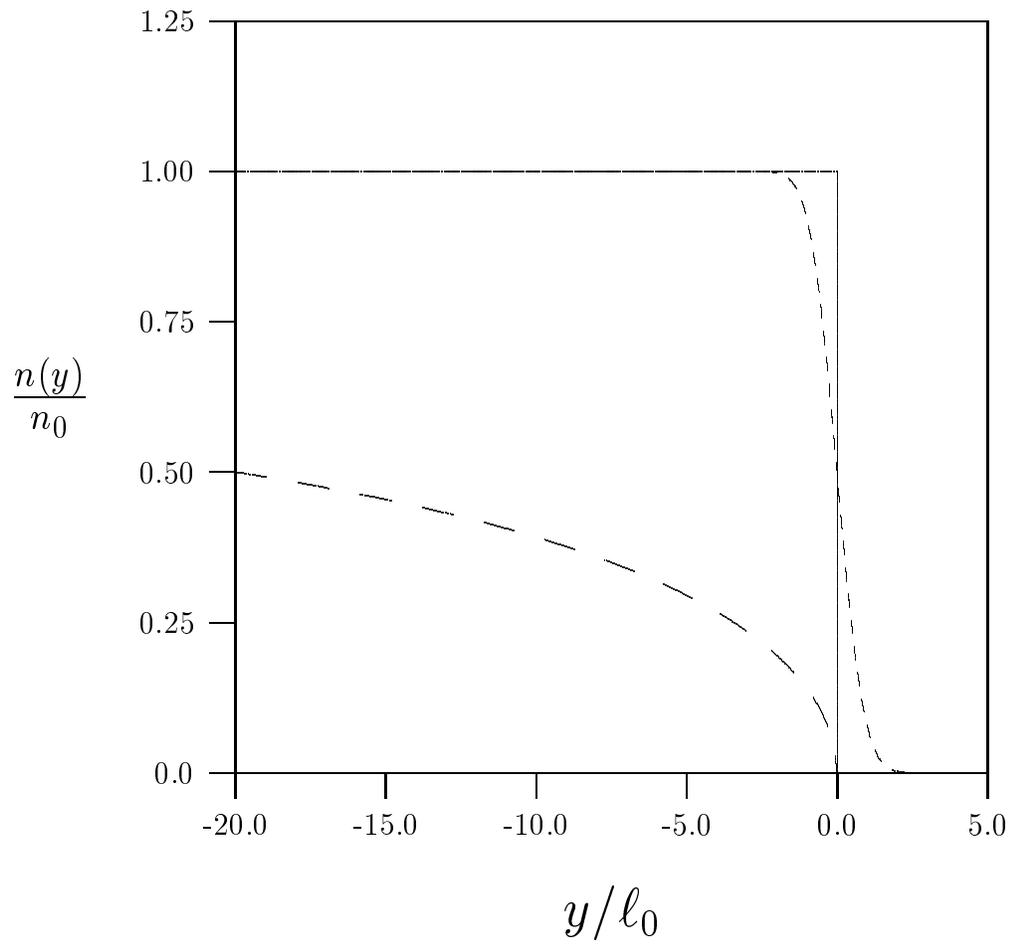
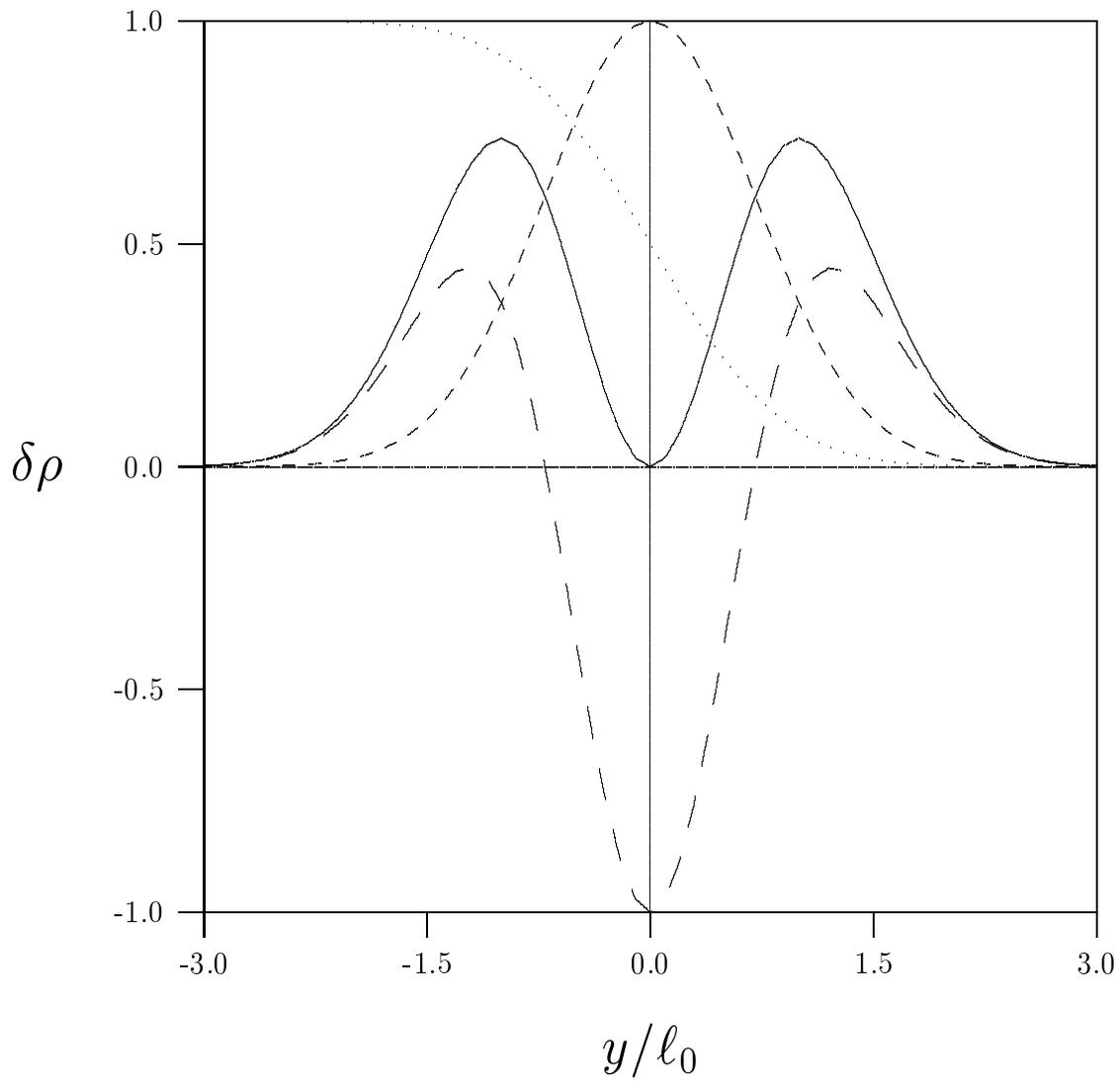
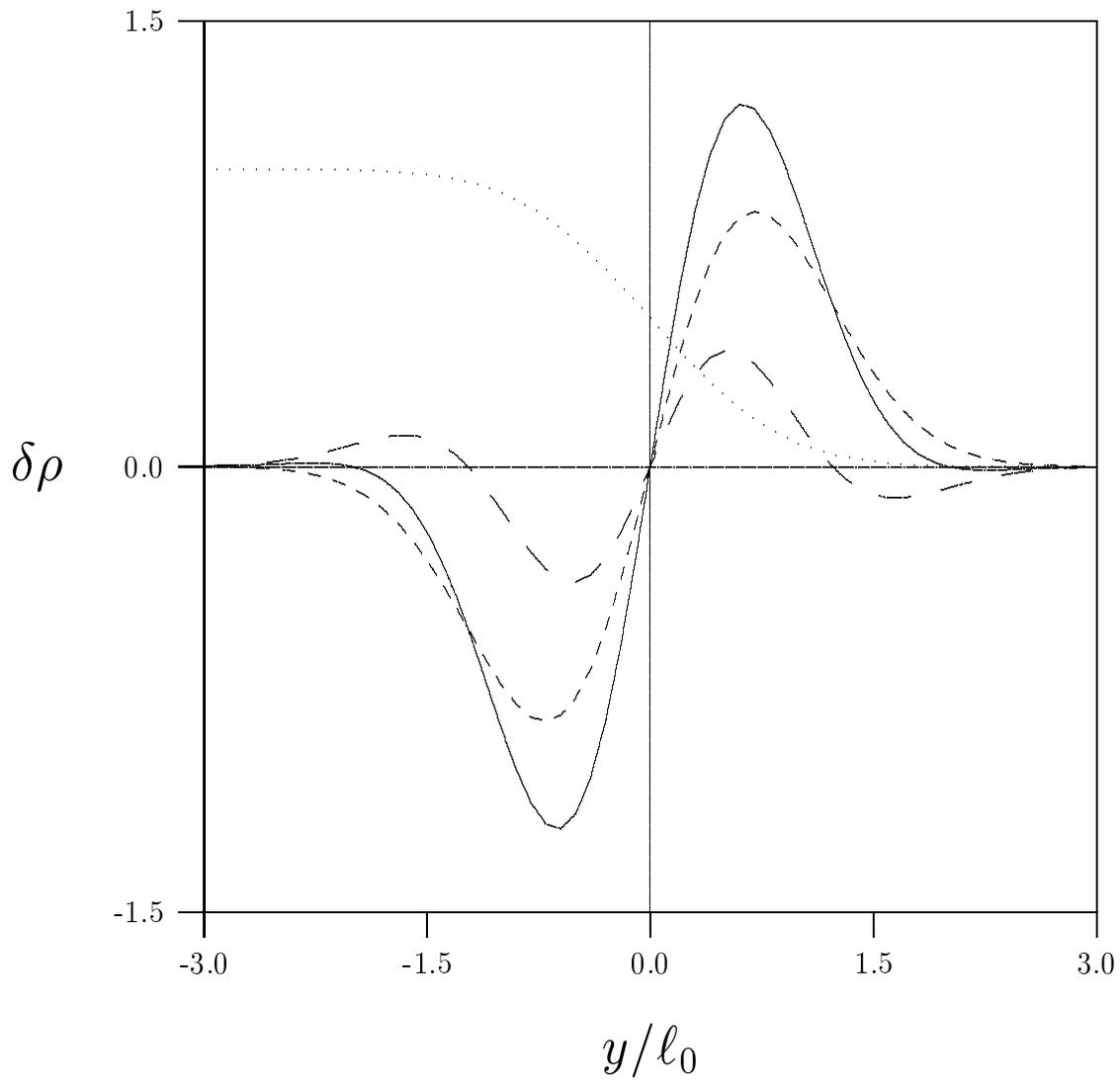
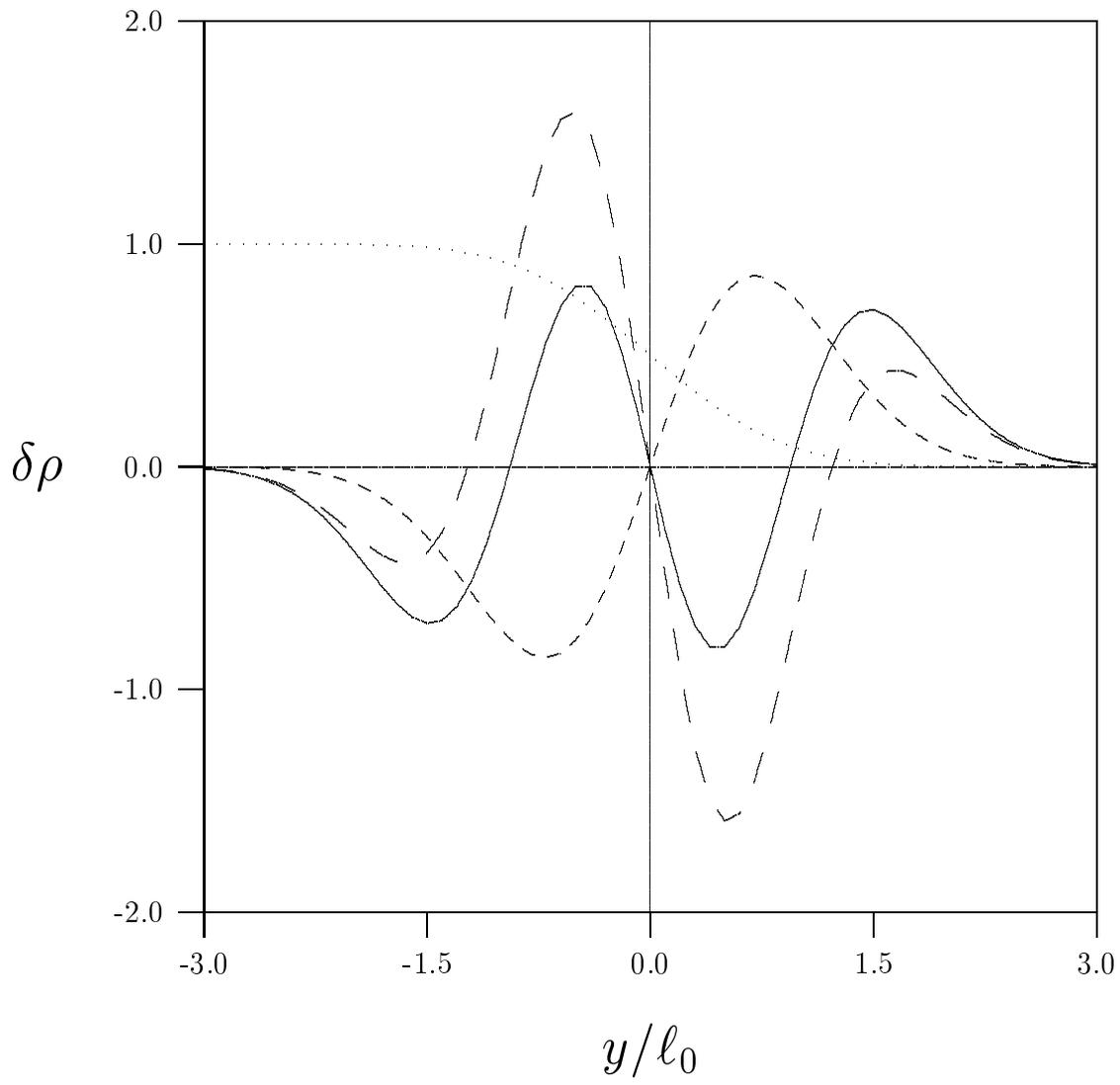
\end{document}